# La prime de risque dans un cadre international : le risque de change est-il apprécié ?

**AROURI Mohamed El Hedi**[*]

**Résumé**

L'objet de cet article est d'étudier les déterminants et la dynamique de la prime de risque des actions dans un cadre international. Pour ce faire, nous utilisons un modèle GARCH multivarié et testons une version conditionnelle du MEDAF international avec déviations de la PPA. Le modèle est estimé sous l'hypothèse d'intégration financière parfaite puis sous l'hypothèse de segmentation partielle conjointement pour cinq marchés : deux marchés développés, deux marchés émergents et le marché mondial. Nos résultats soutiennent le MEDAFI et indiquent que le risque des taux de change est rémunéré internationalement. Cependant, nous trouvons que la prime de change varie considérablement dans le temps et d'un marché à l'autre.

**Abstract**

In this article, we investigate whether exchange rate risk is priced. We use a multivariate GARCH-in-Mean specification and test alternative conditional international CAPM versions. Our results support strongly the international asset-pricing model that includes exchange rate risk for both developed and emerging stock markets. However, there are important time and cross-country variations in the relative size and dynamics of different risk premia.

**Mots clefs** : MEDAFI, Intégration, Risque de change.
**Keywords**: ICAPM, Integration, Currency Risk.
**JEL** : G12, F31, C32.

[*] EconomiX, Université Paris X-Nanterre
Bât G, 200, avenue de la République 92001 Nanterre France.
Tél. +33.1.40.97.77.90.   E-mail : arourix@yahoo.fr
L'auteur tient remercier G. Prat, C. Harvey et M. Raffaëlli pour leurs remarques et suggestions utiles.

# 1- Introduction

La théorie financière nous apprend que lorsqu'il investit dans un actif risqué, l'investisseur est incertain de sa rentabilité, et, s'il est adverse au risque, il exige une prime de risque. Cette prime de risque joue un rôle crucial dans les décisions d'achat et de vente des titres financiers et donc dans toute stratégie de gestion de portefeuille. La mesure de cette prime de risque nécessite la conception d'un modèle d'évaluation des actifs financiers permettant la distinction des sources systématiques et spécifiques de risque. A cette fin, le modèle d'évaluation des actifs financiers (MEDAF) de Sharpe (1964) et Lintner (1965) a été très longtemps utilisé dans le cas des marchés strictement segmentés.

En supposant que les distributions des prix des actifs financiers sont conformes au concept de marché de capitaux parfaitement intégré, de nombreuses extensions internationales du modèle (MEDAFI) furent présentées. La première génération du MEDAFI repose sur l'hypothèse que les investisseurs, indépendamment de leur nationalité, utilisent le même indice des prix pour déflater les rentabilités des différents actifs financiers. Ces modèles constituent des transpositions en termes nominaux du MEDAF domestique. Dans ce contexte, le portefeuille optimal de tout investisseur est une combinaison du portefeuille de marché mondial et de l'actif sans risque, (voir, entre autres, Grauer et al. (1976)). Toutefois, si la parité des pouvoirs d'achat (PPA) n'est pas vérifiée, les investisseurs des différents pays se trouvent face à des opportunités de consommation différentes. Ceci implique une hétérogénéité dans leur appréciation de la rentabilité réelle du même actif financier. Des versions plus « internationales » du modèle furent alors proposées par Solnik (1974), Sercu (1980), Stulz (1981) et Adler et Dumas (1983). Par opposition aux modèles précédents, ces modèles de deuxième génération sont plus généraux et intègrent les problèmes liés aux déviations de la PPA. Enfin, il importe de souligner que les études empiriques montrent que les déviations de la PPA sont la règle plutôt que l'exception. Cela dit, il semble plus raisonnable d'intégrer les déviations de la PPA dans les modèles internationaux d'évaluation des actifs financiers.

L'objet de cet article est d'étudier les déterminants et la dynamique de la prime internationale de risque conjointement pour des marchés émergents et développés. L'article s'intéresse notamment à la contribution de l'éventuelle prime de change dans la prime de risque totale. Il propose en particulier de dépasser certaines limites des travaux antérieurs. En effet, dans la littérature empirique des marchés financiers, de nombreux travaux ont tenté d'étudier la prime internationale de risque. Si les résultats des premiers travaux basés sur des approches statiques sont très hétérogènes, (voir par exemple Solnik (1974) et Korajczyk et Viallet (1989)), récemment des travaux utilisant des modélisations conditionnelles ont montré que le risque des taux de change est internationalement rémunéré du-moins pour les marchés des pays développés. Toutefois, les résultats de ces études peuvent s'avérer difficilement généralisables aux cas des pays émergents en raison de leurs hypothèses de base.

En supposant que les marchés boursiers sont parfaitement intégrés, que la PPA n'est pas vérifiée et que l'inflation locale est non-stochastique, Dumas et Solnik (1995) utilisent la méthode des moments généralisés (MMG) pour tester une version conditionnelle du MEDAFI. Cette approche présente notamment l'avantage de permettre aux primes de risque de varier au cours du temps. Les auteurs montrent que la prime de change est significative pour les quatre plus grands marchés financiers : l'Allemagne, le Royaume-Uni, les Etats-Unis et le Japon. Toutefois, la méthode MMG employée dans cette étude ne permet pas de spécifier la dynamique des seconds moments et donc de mesurer l'importance de la prime de change dans la prime de risque totale.

Afin de surmonter cette difficulté, De Santis et Gérard (1998) utilisent l'approche GARCH multivarié. Contrairement à la méthode MMG, cette approche permet de spécifier la dynamique des seconds moments conditionnels et donc d'évaluer la contribution de chaque facteur de risque à la formation de la prime totale de risque. De Santis et Gérard (1998) font les mêmes hypothèses que Dumas et Solnik (1995) et étudient les mêmes marchés. Les auteurs montrent que la contribution du risque de change dans la prime totale ne peut pas être détectée en se basant sur des approches statiques. Au contraire, le recours aux approches dynamiques avec des prix et des quantités de risque variables dans le temps montre que la prime de change est statistiquement et économiquement significative pour les marchés financiers des pays développés étudiés. En travaillant sous les mêmes hypothèses, Carrieri (2001), Hardouvelis et al. (2002), De Santis et al. (2003) et Cappiello et al. (2003) confirment ces résultats pour d'autres pays européens.

Au cours des dernières années, quelques travaux empiriques ont essayé de généraliser ces études aux marchés des pays émergents. Ces travaux se distinguent les uns des autres par leur degré d'adaptabilité aux particularités des marchés des pays émergents. Tai (2004) applique un processus GARCH multivarié et teste une version conditionnelle internationale du MEDAF où la PPA n'est pas vérifiée. L'étude porte sur des marchés émergents asiatiques (Hong Kong, Singapour, Malaisie). Les résultats de cette étude montrent que le risque de change est rémunéré. Toutefois, le modèle de Tai (2004) suppose, à l'instar des études citées plus-haut, que l'inflation locale est non-aléatoire et que les marchés étudiés sont parfaitement intégrés. Ces hypothèses peuvent paraître fortes dans le cas des marchés des pays émergents souvent marqués par une segmentation financière partielle et par des taux d'inflation volatiles, (voir Bekaert et Harvey (1995) et Karolyi et Stulz (2002)).

Gérard et al. (2003) étudient un MEDAFI à segmentation partielle où les primes de risque sont déterminées par une combinaison de facteurs internationaux et nationaux de risque. Les auteurs trouvent que le risque des taux de change est apprécié internationalement pour quelques marchés émergents. Toutefois comme le signalent les auteurs eux-mêmes, les conclusions de cette étude doivent être considérées avec précaution. En effet, le risque de change n'est pas spécifié explicitement dans le modèle comme un facteur propre de risque. Plus récemment, Phylaktis et Ravazzolo (2004) proposent un modèle MEDAFI à deux régimes qui spécifie explicitement le risque de change comme facteur de risque. Ce modèle considère que les marchés sont strictement segmentés dans un premier temps et deviennent parfaitement intégrés dans un second temps. L'étude porte sur des marchés émergents asiatiques (Corée, Malaisie, Taiwan, Thaïlande, Indonésie et les Philippines) et applique la méthode GARCH (1,1) multivarié. Les auteurs trouvent que le risque de change est apprécié et que le modèle avec déviations de la PPA est économétriquement meilleur que le modèle sans risque de change. Cependant, le modèle de Phylaktis et Ravazzolo (2004) n'envisage que les deux cas extrêmes (segmentation stricte et intégration parfaite), alors que les études les plus récentes montrent que la plupart des marchés émergents sont partiellement segmentés.

Cet article se distingue des travaux antérieurs par trois points principaux. D'abord, nous n'imposons pas de restrictions *a priori* sur les distributions des taux d'inflation. Précisément, nous faisons recours aux taux de change réels, ce qui permettra de surmonter certaines difficultés associées à l'estimation du modèle MEDAFI avec déviations de la PPA dans le cas des pays dont les taux d'inflation sont volatiles. Ensuite, nous estimons le modèle proposé sous l'hypothèse d'intégration financière parfaite, puis sous l'hypothèse plus générale de segmentation partielle. Enfin, en utilisant la méthodologie GARCH multivarié, notre investigation empirique porte en même temps sur des marchés développés et émergents.

La deuxième section de l'article présente une version conditionnelle du MEDAF international avec déviations de la PPA, la troisième section expose la méthodologie retenue pour estimer ce modèle, la quatrième section présente les données et quelques analyses préliminaires, la cinquième section expose les résultats empiriques et la sixième section en tire les principales conclusions.

## 2- Un modèle international d'évaluation des actifs financiers

Dans cette section, nous présentons une version conditionnelle du modèle international d'évaluation des actifs financiers qui tient compte des déviations de la parité des pouvoirs d'achat et d'une éventuelle segmentation financière partielle. Pour commencer, nous reprenons le MEDAFI à intégration financière parfaite proposé par Adler et Dumas (1983).

Considérons un univers avec $L+1$ pays et monnaies et $N = n+L+1$ actifs : $n$ actifs risqués, $L$ actifs sans risque de marché (un de chaque pays hormis le pays de la monnaie de référence (pays $L+1$)) et l'actif sans risque du pays de la monnaie de référence. Supposons en plus que les marchés nationaux sont parfaitement intégrés et que la PPA n'est pas vérifiée.

Désignons par $P_i^c$ le prix de l'actif $i$ mesuré dans la monnaie du pays de référence $c$, $R_i^c$ la rentabilité nominale de cet actif exprimée dans la monnaie de référence, $E(R_i^c)$ et $\sigma_i^c$ ses premier et second moments instantanés. $R_i^c$ est supposé suivre le processus brownien suivant :

$$R_i^c dt \equiv \frac{dP_i^c}{P_i^c} = E(R_i^c)dt + \sigma_i^c dz_i^c \quad ; \quad i = 1,2,...,n \qquad (1)$$

où la variable $z_i^c$ suit un processus de Wiener standard.

Le taux d'inflation dans chaque pays exprimé dans la monnaie de référence est supposé suivre le même processus. Formellement, désignons par $I_k^c$ l'indice général des prix dans le pays $k$ mesuré dans la monnaie de référence, $\pi_k^c$ le taux d'inflation exprimé dans la monnaie de référence et $E(\pi_k^c)$ et $\sigma_{\pi k}^c$ ses premier et second moments instantanés. Donc, $\pi_k^c$ suit le processus suivant:

$$\pi_k^c dt \equiv \frac{dI_k^c}{I_k^c} = E(\pi_k^c)dt + \sigma_{\pi k}^c dz_{\pi k}^c \quad ; \quad k = 1,2,...,L+1 \qquad (2)$$

où la variable $z_{\pi k}^c$ suit un processus de Wiener standard.

Notons que $\pi_k^c$ est aléatoire dans la mesure où ses fluctuations reflètent les fluctuations de l'inflation locale exprimée en monnaie locale du pays $k$, $\pi_k$, et celles du taux de change entre la monnaie du pays $k$ et la monnaie du pays de référence $c$.

L'investisseur international cherche à maximiser l'utilité espérée de sa consommation réelle future. Son programme peut être écrit ainsi :

$$\underset{C_k, \varpi_k}{Max} \int_t^T U(C_k, I_k, \tau) d\tau$$

$$\text{s.c.} \quad dW_k^c = \left[\sum_{i=1}^n \varpi_i \left(E(R_i^c) - R_f^c\right) + R_f^c\right] W_k^c dt - C_k dt + W_k^c \sum_{i=1}^n \varpi_i \sigma_i^c dz_i^c \quad (3)$$

où $C_k$ est le flux de consommation nominale, $W_k^c$ est le niveau de la richesse nominale du pays $k$ mesuré dans la monnaie de référence, $\varpi_i$ est la part de la richesse investie dans le titre $i$, et $U(C_k, I_k, \tau)$ est une fonction d'utilité satisfaisant les hypothèses traditionnelles, en particulier, l'homogénéité de degré zéro.

La résolution de ce problème d'optimisation permet de déduire l'allocation optimale de portefeuille, et donc, la fonction de demande des actifs par chaque investisseur international. Etant donné l'objectif de l'article, nous nous intéressons à la prime de risque que demande chaque investisseur international pour investir dans un actif risqué. Supposons que l'offre est exogène et que les marchés sont en équilibre (*i.e.* pour chaque actif l'offre est égale à la demande), et dérivons, par agrégation des demandes individuelles des investisseurs internationaux, l'expression suivante:[1]

$$E(R_i^c) - R_f^c = \delta_m Cov(R_i^c, R_m^c) + \sum_{k=1}^L \delta_k Cov(R_i^c, \pi_k^c) \quad \forall i \quad (4)$$

où $R_m^c$ est l'excès de rentabilités nominales du portefeuille du marché mondial. $\delta_m$ et $\delta_k$ sont définis comme il suit:

$$\delta_m = \theta_m = \frac{1}{\sum_{k=1}^L \frac{W_k^c}{W^c} \times \frac{1}{\theta_k}} \quad \text{et} \quad \delta_k = \theta_m \left(\frac{1}{\theta_k} - 1\right) \frac{W_k^c}{W^c}$$

avec $\theta_k$ est le coefficient d'aversion relative au risque du pays $k$ et $\theta_m$ est la moyenne des aversions nationales au risque pondérées par les parts relatives des richesses nationales dans la richesse totale.

Notons que dans le MEDAF traditionnel de Sharpe (1964) et Lintner (1965), la covariance entre la rentabilité du titre $i$ et celle du portefeuille de marché représente le risque de marché du titre $i$, donc $\delta_m$ est interprété comme le prix de risque du marché mondial. En outre, les covariances entre la rentabilité de l'actif $i$ et le taux d'inflation de chacun des pays mesurent les sources additionnelles de risque induites par les déviations de la PPA. Précisément, le terme $Cov(R_i^c, \pi_k^c)$ mesure l'exposition de l'actif $i$ à l'inflation locale et au risque de change associés au pays $k$. En effet, les variations de $\pi_k^c \cong \pi_k + dS_k/S_k$ sont dues aux variations de l'inflation locale exprimée en monnaie locale, $\pi_k$, et aux variations des taux de change entre la monnaie du pays $k$ et la monnaie de référence $c$.

Dans les tentatives de validation empirique du modèle décrit par la relation (4), l'inflation exprimée en monnaie locale est toujours supposée non-stochastique. Cette hypothèse est plausible dans le cas des marchés des pays développés. En effet, dans ces marchés développés, les taux d'inflation sont négligeables par rapport aux variations des taux de change. Sous cette hypothèse, le terme $Cov(R_i^c, \pi_k^c)$ mesure l'exposition de l'actif *i* au risque de change du pays *k*. $\delta_k$ est alors interprété comme le prix du risque de change associé au pays *k*.

---
[1] Pour une présentation plus détaillée de la résolution de ce programme voir Adler et Dumas (1983).

Cependant, pour de nombreux marchés émergents, l'inflation exprimée en monnaie locale est très volatile. Il n'est donc pas judicieux de substituer les variations du taux de change nominal au taux d'inflation exprimée en monnaie de référence, $\pi_k^c$. Par ailleurs, l'utilisation du risque de taux de change nominal afin d'approcher l'inflation exprimée en monnaie de référence pourrait causer des défauts de spécification au niveau de l'estimation de la prime de risque puisqu'elle ne tient pas compte de l'ajustement de l'inflation domestique exprimée en monnaie locale.

Toutefois, supposer que l'inflation locale est aléatoire complique l'estimation du modèle. En effet dans ce cas, nous avons deux termes de covariance pour chaque taux de change. Ce double problème de mesure empirique augmente grandement le nombre de paramètres à estimer et complique l'estimation du modèle conjointement pour plusieurs marchés. Afin de surmonter cette difficulté, nous proposons dans cet article d'approcher le taux d'inflation exprimée en monnaie de référence, $\pi_k^c$, par les variations du taux de change réel. En ce faisant, nous aurons besoin uniquement de supposer que l'inflation locale du pays de référence $\pi_c$ est non-aléatoire. En supposant que $\pi_c$ est non-aléatoire, $\pi_k^c$ peut être assimilée aux variations du taux de change réel entre la monnaie du pays *k* et la monnaie du pays de référence *c*.

Pour montrer cela, désignons par $S_k^r$ le taux de change réel du pays k par rapport au pays de référence c. Donc: $S_k^r = S_k \times I_k / I_c$, où $I_k$ est le niveau général des prix dans le pays k, $I_c$ est le niveau général des prix dans le pays de référence c et $S_k$ le taux de change nominal mesurant le prix en monnaie locale d'une unité de la monnaie du pays de référence. Nous pouvons réécrire la relation précédente ainsi : $S_k^r \times I_c = S_k \times I_k$, ou encore $I_k^c = S_k^r \times I_c$ avec $I_k^c = I_k \times S_k$ est le niveau général des prix dans le pays k exprimé en monnaie du pays de référence c. Le taux d'inflation du pays k exprimé en monnaie de référence, $\pi_k^c$, est alors donné par:

$$\pi_k^c = \Delta \ln(I_k^c) = \Delta \ln(I_c) + \Delta \ln(S_k^r) = \pi_c + \Delta \ln(S_k^r) \tag{5}$$

Donc en supposant que l'inflation dans le pays de référence $\pi_c$ est non-aléatoire, nous pouvons approcher l'inflation locale exprimée en monnaie de référence, $\pi_k^c$, par les variations du taux de change réel. Economiquement, cette approximation montre que le risque des déviations de la PPA peut être approché par les variations du taux de change réel puisque ces dernières proviennent de l'effet combiné des variations du différentiel d'inflation entre le pays *k* et le pays de référence *c* et des variations de la valeur nominale de la monnaie.[2] Cela dit, en intégrant (5) dans (4) et en supposant que les distributions des rentabilités financières sont variables au cours du temps, nous obtenons la version suivante du modèle :

$$E(R_{i,t}^c / \Omega_{t-1}) - R_{f,t}^c = \delta_{m,t-1} Cov(R_{i,t}^c, R_{m,t}^c / \Omega_{t-1}) + \sum_{k=1}^{L} \delta_{k,t-1} Cov(R_{i,t}^c, R_{k,t}^c / \Omega_{t-1}) \quad \forall i \tag{6}$$

---

[2] L'utilisation des taux de change réels permet aussi de surmonter certaines difficultés liées aux régimes de change fixes, par exemple les dévaluations fréquentes dans les pays émergents.

où $R_{k,t}^c$ mesure la rentabilité du taux de change réel du pays *k*, $\Omega_{t-1}$ est le vecteur de variables d'information disponibles à la fin de la période (t-1) et où seule l'inflation dans le pays de la monnaie de référence $\pi_c$ est supposée non-aléatoire.

Enfin, notons que le modèle décrit par la relation (6) suppose que les marchés financiers sont parfaitement intégrés. Or, si l'on en croit les résultats des études les plus récentes, les marchés financiers, notamment ceux des pays en développement, sont éventuellement caractérisés par un certain degré de segmentation, (voir Karolyi et Stulz (2002)). Autrement dit, les rentabilités des actifs financiers dans ces marchés dépendent d'une combinaison de facteurs globaux et locaux de risque. Cela dit, nous étendons le modèle décrit par la relation (6) au cas plus général d'intégration financière partielle.

Dans le cadre des marchés partiellement segmentés, les rentabilités boursières sont déterminées par deux sources de risque : le risque du marché global et le risque domestique résiduel, (voir par exemple Gérard et al. (2003)). Le risque domestique résiduel non-corrélé au portefeuille du marché mondial est mesuré ainsi:

$$Var(\text{Res}_{it}^c / \Omega_{t-1}) = Var(R_{it}^c / \Omega_{t-1}) - Cov(R_{it}^c, R_{mt}^c / \Omega_{t-1})^2 / Var(R_{mt}^c / \Omega_{t-1})$$

Une version conditionnelle du MEDAFI à segmentation partielle avec déviations de la PPA peut être formalisée comme il suit:

$$E(R_{i,t}^c / \Omega_{t-1}) - R_{f,t}^c = \delta_{m,t-1} Cov(R_{i,t}^c, R_{m,t}^c / \Omega_{t-1}) + \sum_{k=1}^{L} \delta_{k,t-1} Cov(R_{i,t}^c, R_{k,t}^c / \Omega_{t-1}) \\ + \delta_{di} Var(\text{Res}_{it}^c / \Omega_{t-1}) \qquad \forall i \qquad (7)$$

où $\delta_{di}$ est le prix du risque domestique du pays *i*. Si les marchés sont parfaitement intégrés, alors le risque domestique n'est pas apprécié et le modèle (7) se réduit au modèle (6).

## 3- Spécification économétrique

Si les marchés sont parfaitement intégrés, le modèle décrit par (6) est valable pour tous les actifs financiers y compris le portefeuille de marché mondial. Sous l'hypothèse d'anticipations rationnelles, pour un monde à *L+1* pays et *n* actifs risqués le système d'équations suivant doit être satisfait à chaque point du temps :

$$r_t = \delta_{m,t-1} h_{m,t} + \sum_{k=1}^{L} \delta_{k,t-1} h_{n+k,t} + \varepsilon_t \quad , \quad \varepsilon_t / \Omega_{t-1} \sim N(0, H_t) \qquad (8)$$

où $r_t$ est le vecteur ($N \times 1$) d'excès de rentabilités exprimés en monnaie de référence *c*, $\varepsilon_t$ est le vecteur ($N \times 1$) des résidus, $H_t$ est la matrice ($N \times N$) des variances-covariances conditionnelles, $h_{n+k,t}$ et $h_{m,t}$ sont respectivement la $n+k^{\text{ème}}$ et la dernière colonne de $H_t$. $h_{n+k,t}$ et $h_{m,t}$ mesurent respectivement les covariances des $N = n+L+1$ actifs risqués avec les *L* taux de change et le portefeuille du marché mondial. Autrement dit, $h_{n+k,t}$ mesure l'exposition au risque des taux de change et $h_{m,t}$ mesure l'exposition au risque du marché mondial.

Dans le système (8), les *n* premières équations sont utilisées pour modéliser les rentabilités des marchés nationaux, les *L* équations suivantes pour modéliser les dynamiques des taux de change et la dernière équation (équation *L+n+1*) pour le portefeuille du marché mondial.

Comme expliqué précédemment, nous étendons le modèle décrit par (8) pour tenir compte d'une éventuelle segmentation partielle des marchés financiers nationaux. Le nouveau système d'équations doit prendre en considération le facteur risque domestique :

$$r_t = \delta_{m,t-1} h_{m,t} + \sum_{k=1}^{L} \delta_{k,t-1} h_{n+k,t} + \delta_d * q_t + \varepsilon_t \quad , \quad \varepsilon_t / \Omega_{t-1} \sim N(0, H_t) \tag{9}$$

où $q_t = D(H_t) - (h_{mt} * h_{mt})/h_{mmt}$ est le vecteur de taille ($N \times 1$) des risques domestiques, $\delta_d$ est le vecteur des prix des risques domestiques, $h_{mt}$ est la $N^{ème}$ colonne de $H_t$ contenant les covariances des actifs avec le portefeuille du marché, $h_{mmt}$ est la variance du portefeuille du marché mondial et $D(H_t)$ est la diagonale de la matrice $H_t$ des variances-covariances.

Les systèmes (8) et (9) impliquent l'estimation simultanée des premiers et seconds moments conditionnels pour les *N* actifs étudiés. Pour ce faire, nous utilisons le modèle GARCH(1,1) multivarié De Santis et Gérard (1997). Formellement :

$$H_t = H_0 * (\tau\tau' - aa' - bb') + aa' * \varepsilon_{t-1}\varepsilon'_{t-1} + bb' * H_{t-1} \tag{10}$$

où $H_0$ est la matrice des variances-covariances non-conditionnelles, *a* et *b* sont deux vecteurs ($N \times 1$) de paramètres inconnus, $\tau$ est le vecteur unitaire de taille ($N \times 1$) et * le produit matriciel de Hadamard (élément par élément). L'extension au cadre multivarié des modèles GARCH(1,1) implique que les termes d'erreur ont une distribution conditionnelle gaussienne de moyenne nulle et de matrice de variances-covariances $H_t$.

En outre, le MEDAF international multifactoriel exposé ci-dessus comporte deux types de risques : le risque de marché et le risque de change. A chaque type de risque, on peut associer une rémunération par unité de risque, (*i.e.* un prix de risque). Considérons en premier lieu la dynamique du prix de risque de marché mondial. Les travaux de Harvey (1991), Bekaert et Harvey (1995) et De Santis et Gérard (1997) montrent que le prix de risque varie dans le temps. De plus, selon Merton (1980) et Adler et Dumas (1983), le prix de risque de marché est une agrégation des aversions pour le risque de tous les investisseurs. Or les investisseurs sont supposés adverses au risque, donc le prix de risque de marché doit être positif à chaque point du temps. A l'instar de De Santis et Gérard (1997,1998), Bekaert et Harvey (1995), De Santis et al. (2003) et Gérard et al. (2003), nous modélisons le prix de risque de marché comme fonction exponentielle de certaines variables d'information liées au cycle économique et financier international : $\delta_{t-1} = \exp(\kappa'_W Z_{t-1})$, où $Z_{t-1}$ est l'ensemble de variables d'information globales observables disponibles en (t-1), $Z_{t-1} \subset \Omega_{t-1}$, et $\kappa_W$ représente les pondérations associées à ces variables d'information globales.

En ce qui concerne le prix de risque des taux de change, la théorie n'impose pas de restriction quant à son signe. Le prix de risque de change peut théoriquement prendre aussi bien des valeurs positives que des valeurs négatives. Pour modéliser les prix des risques des taux de change, nous utilisons à l'instar de De Santis et al. (2003) et Phylaktis et Ravazzolo (2004) des fonctions linéaires. Ainsi, les prix des risques de change sont supposés varier linéairement en fonction des variables instrumentales: $\delta_{k,t-1} = \kappa_k^{'} Z_{t-1}$, où $\kappa_k$ représente les pondérations associées à ces variables.[3]

Sous l'hypothèse d'une distribution conditionnelle multivariée normale, la fonction de vraisemblance peut être écrite comme il suit :

$$\ln L(\psi) = -\frac{TN}{2}\ln(2\pi) - \frac{1}{2}\sum_{t=1}^{T}\ln|H_t(\psi)| - \frac{1}{2}\sum_{t=1}^{T}\varepsilon_t'(\psi)H_t^{-1}(\psi)\varepsilon_t(\psi) \qquad (11)$$

où $\psi$ est le vecteur des paramètres inconnus et $T$ est le nombre d'observations. Dans la mesure où l'hypothèse de normalité est souvent rejetée dans le cas des séries boursières, nous utilisons la méthode du quasi-maximum de vraisemblance (QMV) de Bollerslev et Wooldridge (1992). Sous certaines conditions de régularité, l'estimateur QMV est valide et asymptotiquement normal. D'abord, l'algorithme de simplex est utilisé pour initialiser le processus. Ensuite, l'estimation du vecteur $\psi$ est réalisée par l'algorithme (BHHH) développé par Berndt, Hall, Hall et Hausman (1974).

## 4- Données et analyse préliminaire

Cette section a deux objectifs : présenter les données que nous utiliserons dans notre investigation empirique et montrer que ces données présentent des caractéristiques modélisables avec des modèles de type GARCH. Trois groupes de données sont à distinguer : les séries de rentabilités, les séries de taux de change et les variables macro-économiques et financières utilisées afin de conditionner les estimations.

Etant donné l'objectif de l'article, l'investigation empirique porte sur les indices des marchés boursiers de quatre pays (deux pays développés (France et Etats-Unis) et deux pays émergents (Singapour et Afrique du Sud)) et l'indice du marché mondial. Ce choix se justifie notamment par le souci de comparer les comportements des prix et des risques dans les marchés des pays développés et des pays émergents. Les observations utilisées sont des cours mensuels de fin de période allant de janvier 1973 à mai 2003, soit 365 observations.

Les cours boursiers des marchés français, américain et singapourien sont issus de *Morgan Stanley Capital International* (MSCI), l'indice du marché boursier sud-africain est obtenu de *Datastream*. L'indice du marché mondial retenu est l'indice MSCI Monde. Les rentabilités boursières sont toutes exprimées en dollar américain, calculées avec réinvestissement des dividendes en excès du taux des eurodollars à 30 jours issu de *Datastream*.

Les taux de change nominaux bilatéraux sont extraits de la *Federal Reserve Bank of St Louis' FRED DataBase* et de *International Financial Statistics* (IFS). Les taux de change réels sont construits en déflatant les taux de change nominaux. Pour ce faire, nous avons utilisé les indices des prix à la consommation (IPC) obtenus de la base de données IFS.

Finalement, le système (8) contient huit équations organisées comme il suit : quatre équations pour les rentabilités boursières des pays étudiés (Etats-Unis, France, Afrique du Sud et Singapour), trois équations pour les variations des séries des taux de change réels (le franc français (TcrFra), le rand sud-africain (TcrAfri) et le dollar singapourien (TcrSinga)) et une équation pour le marché mondial. Il s'ensuit que chacune des huit équations du MEDAF international multifactoriel multivarié représenté par le système (8) contient quatre primes de risque : une associée au risque du portefeuille du marché mondial et trois liées aux risques des taux de change (une prime pour chacune des monnaies considérées). Dans le cas plus général des marchés financiers nationaux partiellement segmentés (système (9)), des primes des risques domestiques seront introduites.

---

[3] Afin de garantir une meilleure comparabilité avec les études précédentes, entre autres Dumas et Solnik (1995), De Santis et Gérard (1998), Carrieri (2001) et De Santis et al. (2003), le même vecteur d'information est utilisé pour modéliser le prix de risque de marché mondial et les prix des risques de change.

Le Tableau 1 résume les statistiques descriptives des séries étudiées. Le panel A présente les rentabilités moyennes ainsi que les tests de normalité et d'autocorrélation. L'Afrique du Sud présente l'excès de rentabilités moyen le plus élevé. Les marchés les plus volatiles sont le marché singapourien et le marché sud-africain. Le marché américain est le marché national le moins volatile. Notons aussi que les séries des taux de change sont moins volatiles que les séries des indices boursiers nationaux.

Les coefficients d'asymétrie, souvent négatifs, indiquent que la distribution des séries est étalée vers la gauche. On signale également le caractère leptokurtique des séries. L'hypothèse de normalité est rejetée pour toutes les séries étudiées, ce qui justifie le choix de la méthode d'estimation du quasi-maximum de vraisemblance.

Le Panel B reporte les corrélations non-conditionnelles des séries de rentabilités et des séries des taux de change. Comme prévu, les corrélations non-conditionnelles des marchés boursiers sont positives. Ces corrélations sont relativement faibles. La corrélation la plus élevée est entre le marché américain et le marché mondial, 85%. Cela peut s'expliquer par la part relativement importante du marché américain dans le marché mondial. La corrélation la plus faible est entre le marché sud-africain et le marché singapourien, 30%. L'Afrique du Sud a la corrélation non-conditionnelle la plus faible avec le marché mondial, 46%. Les corrélations entre les marchés émergents et les marchés développés étudiés sont plus faibles que celles entre les marchés développés, ce qui montre l'intérêt de ces marchés en matière de diversification internationale des portefeuilles. Les corrélations entre les séries de change et les séries boursières sont faibles et souvent négatives. Le test de Ljung-Box d'ordre 12 montre l'absence d'autocorrélation sérielle pour toutes les séries boursières. Ce résultat semble être confirmé par les autocorrélations des excès de rentabilités présentées dans le panel C.

Le panel D présente les autocorrélations des carrés des séries de rentabilités boursières et des taux de change réels étudiées et le panel E expose les corrélations croisées des carrés de ces séries avec le portefeuille de marché mondial. Pour la plupart des séries étudiées, seules les autocorrélations d'ordre un sont significatives, ce qui pourrait aller en faveur d'une modélisation GARCH d'ordre 1. En outre, sauf exception, seules les corrélations croisées instantanées sont significatives. Quand on analyse les corrélations croisées d'ordre {-2,-1,1,2}, seulement 11 corrélations sur un total de 112 corrélations sont significatives. Ce résultat laisse penser qu'au moins pour nos séries de rentabilités mensuelles, les interdépendances en terme de volatilité ne sont pas importantes.

En ce qui concerne le choix des variables d'information à utiliser pour conditionner l'estimation du prix du risque de marché mondial et des prix des risques de change, nous nous inspirons principalement des résultats des travaux antérieurs en finance internationale, (voir notamment Harvey (1991), Ferson et Harvey (1993), De Santis et Gérard (1997,1998), Bekaert et Harvey (1995), De Santis et al. (2003) et Gérard et al. (2003)). Les variables retenues sont censées refléter les informations concernant le cycle financier et économique mondial dont disposent les investisseurs à la date (t-1), (voir Harvey (1991) et Dumas (1994)). Ainsi, le vecteur $Z_{t-1}$ de variables d'information internationales contient: un terme constant, le rendement en dividende (*dividend price ratio*) du portefeuille de marché mondial en excès du taux des eurodollars à 30 jours (RDM), la variation mensuelle d'une prime de terme américaine (DPTEU), une prime de défaut américaine (PDEU) et la variation mensuelle du rendement d'un certificat américain de trésorerie à 30 jours (DTIM).[4] La prime de terme est mesurée par la différence entre un taux d'intérêt court (un certificat de trésorerie américain à 3 mois) et un taux long (un bon de trésor américain à 10 ans) et la prime de défaut est mesurée par l'écart de rendements entre une obligation notée Baa par l'agence Moody's et une obligation notée Aaa. Toutes ces variables d'information proviennent de MSCI et de IFS et sont utilisées avec un retard par rapport aux séries de rentabilités.

Les statistiques descriptives de ces variables sont résumées dans le Tableau 2. Les corrélations entre les variables globales d'information sont relativement faibles, ce qui suggère que le vecteur d'information $Z$ ne contienne pas d'informations redondantes.

## 5- Résultats empiriques

Comme nous l'avons explicité ci-dessus nous allons d'abord estimer le modèle sous l'hypothèse d'intégration financière parfaite. Ensuite, le modèle sera re-estimé sous l'hypothèse plus réaliste des marchés financiers nationaux partiellement segmentés.

### 5.1- Cas des marchés boursiers parfaitement intégrés

Dans cette sous-section, nous supposons que les marchés financiers étudiés sont parfaitement intégrés dans le marché mondial. Le système décrit par les équations (8) et (10) constitue notre modèle de base.

*5.1.1- Résultats et tests de spécification*

Le Tableau 3 reporte les résultats de l'estimation du modèle par la méthode du quasi-maximum de vraisemblance (QMV) ainsi que quelques tests de spécification. Le Panel A présente les paramètres relatifs à la dynamique du prix de risque du marché mondial et des prix des risques de change des monnaies étudiées vis-à-vis du dollar américain. Le prix du risque du portefeuille de marché mondial est déterminé par le terme constant, le rendement en dividende, la variation mensuelle de la prime de terme, la prime de défaut et la variation mensuelle du rendement du certificat américain de trésorerie à 30 jours. Les prix des risques de change sont notamment déterminés par la variation mensuelle de la prime de terme et la variation mensuelle du rendement du certificat américain de trésorerie à 30 jours.

Le panel B du Tableau 3 présente la structure des seconds moments conditionnels. Les coefficients *a* et *b* sont significatifs pour toutes les séries de rentabilités étudiées. Les valeurs estimées du vecteur *b* (qui relient les seconds moments à leurs valeurs passées) sont largement supérieures à celles du vecteur *a* (qui relient les seconds moments aux innovations passées), ce qui témoigne de changements graduels dans la dynamique de la volatilité conditionnelle. En outre, certains marchés manifestent une forte persistance. Ces résultats sont en accord avec les études antérieures utilisant des spécifications GARCH.

---

[4] Harvey (1991) montre que la prime de terme américaine présente une corrélation de 87% avec la prime de terme pondérée mondiale.

Le panel C présente quelques tests d'hypothèses jointes liées aux dynamiques du prix de risque du marché mondial et des prix des risques de change. Ces hypothèses sont testées par le test robuste de Wald calculé à partir de l'estimation du modèle par la méthode du quasi-maximum de vraisemblance.

L'hypothèse de base selon laquelle le prix du risque mondial est constant est rejetée à tous les seuils conventionnels de significativité. Le prix du risque mondial est variable au cours du temps en fonction des variables reflétant la conjoncture économique et financière internationale. Ce résultat est en accord avec les études les plus récentes en finance internationale, (voir par exemple Gérard et al. (2003), Phylaktis et al. (2004) et Tai (2004)).

La Figure 1.a (voir annexe) montre l'évolution du prix de risque du marché mondial. Le prix de risque estimé est très volatile. Nous nous intéressons à la tendance générale plutôt qu'aux fluctuations à court terme. Ceci dit, nous reportons aussi la série filtrée par la méthodologie de Hodrick et Prescott (1996). Le filtre HP permet de séparer les mouvements de court terme (cycles) du mouvement de long terme (tendance). La série filtrée atteint ses valeurs les plus élevées dans les années 1970, se réduit dans les années 1980 puis s'accroît à nouveau dans la première partie des années 1990. Ceci confirme les résultats de De Santis et Gérard (1997). De 1993 à 2000, la série filtrée du prix du risque de marché est en-dessous de sa moyenne sur la période entière 1973-2003. Enfin, le prix du risque de covariance du marché mondial s'accroît rapidement à partir de 2001, traduisant l'incertitude que traversent les marchés financiers internationaux dans les dernières années.

En ce qui concerne les prix des risques des taux de change, nous commençons par tester les hypothèses selon lesquelles ces prix sont individuellement puis conjointement non significativement différents de zéro. Si ces prix s'avèrent statistiquement significatifs, nous testerons s'ils sont constants ou variables suivant les dates.

Les hypothèses selon lesquelles les prix des risques de change sont individuellement nuls sont rejetées pour tous les pays étudiés à tous les seuils conventionnels. Les hypothèses selon lesquelles les prix des risques de change sont constants sont rejetées à tous les niveaux de significativité pour le franc français et le dollar singapourien et à 10% pour l'Afrique du Sud. Ces résultats sont confirmés par les tests d'hypothèses jointes. L'hypothèse selon laquelle les prix des risques des taux de change sont conjointement égaux à zéro et l'hypothèse selon laquelle les prix des risques de change sont conjointement constants sont rejetées à tous les seuils habituels de significativité. Carrieri et al. (2005) testent en deux étapes une version du MEDAFI incluant, en plus du risque du marché mondial, des risques des taux de change composites réels et trouvent des résultats similaires.

Le Panel D présente quelques tests sur les résidus. L'hypothèse de normalité est rejetée pour tous les marchés étudiés. Il y a cependant lieu de signaler que les coefficients d'asymétrie (skewness) et d'aplatissement centré (kurtosis) sont plus faibles que ceux des séries de rentabilités présentées dans le Tableau 1. En outre, le test Ljung-Box d'absence d'autocorrélation d'ordre 12 a été appliqué sur les résidus. A l'exception des séries des taux de change, les résultats de ce test ne permettent pas de rejeter l'hypothèse nulle d'absence d'autocorrélation.

Pour résumer, les résultats de nos estimations montrent que le risque de change est rémunéré internationalement et que sa rémunération est variable au cours du temps aussi bien pour les marchés développés que pour les marchés émergents étudiés. Ceci a des implications importantes en matière de gestion internationalisée des portefeuilles et en matière d'évaluation internationale des actifs financiers. Une appréciation de l'importance et de la dynamique des primes des risques de change s'avère alors nécessaire. Heureusement, l'approche économétrique utilisée dans ce travail permet la décomposition de la prime de risque totale en fonction des différents facteurs de risque. La méthodologie retenue permet aussi d'étudier les dynamiques des différentes composantes de la prime totale. Les résultats de cette analyse seront présentés en deçà.

*5.1.2- Analyse des primes de risque*

Une fois que les prix des risques et les seconds moments conditionnels sont estimés, il est relativement facile d'analyser la dynamique des primes associées aux différents facteurs de risque. Dans cet article, la variation dans le temps de la prime de risque a deux sources distinctes puisque les prix des risques et les seconds moments sont autorisés à varier au cours du temps. Les Figures de 2 à 6 présentent l'évolution des primes de risque calculées pour les marchés nationaux et le marché mondial. Les primes associées à chacun des facteurs de risque sont mesurées comme il suit :

Prime de Risque de Marché Mondial (PRM): $PRM_{it} = \delta_{m,t-1} Cov(R_{it}, R_{mt} / Z_{t-1})$

Prime de Risque de Change (PRC): $PRC_{it} = \sum_{k=1}^{3} \delta_{k,t-1} Cov(R_{it}, R_{4+k,t} / Z_{t-1})$

Prime Totale (PT): $PT_{it} = \delta_{m,t-1} Cov(R_{it}, R_{mt} / Z_{t-1}) + \sum_{k=1}^{3} \delta_{k,t-1} Cov(R_{it}, R_{4+k,t} / Z_{t-1})$

Pour des raisons de clarté, les primes de risque associées au facteur risque du portefeuille du marché mondial ne sont pas représentées. La prime de marché s'obtient en faisant la différence entre la prime totale et la prime de risque de change. Notons aussi que le risque de change provient des expositions agrégées aux fluctuations des taux de change des différents investisseurs dans les différents pays consommant des biens exprimés dans leurs monnaies respectives. Donc, une prime de change existe aussi pour l'investisseur représentatif américain même si toutes les rentabilités sont exprimées en dollar américain.

Pour les Etats-Unis et le marché mondial, les évolutions des primes de risque montrent que la prime de marché est la composante la plus significative de la prime totale. Pour les autres marchés, la prime associée au risque des taux de change contribue plus significativement à la formation de la prime de risque totale. Néanmoins, le poids et le signe de la prime de risque de change varient considérablement dans le temps et d'un marché à l'autre. En moyenne, les primes associées au risque des taux de change sont négatives, ce qui montre l'importance que les investisseurs globaux accordent à la couverture du risque de change. En d'autres termes, les investisseurs sont prêts à payer une partie de leur prime totale pour se protéger contre le risque des taux de change.

Le contenu informationnel des primes de risque totales est riche. En effet, ces dernières nous informent sur les événements qui ont induit des changements substantiels dans les primes de risque que les investisseurs globaux exigent. Les investisseurs représentatifs supposés adverses au risque exigent des primes de risque plus élevées en période de fortes incertitudes. Pour tous les marchés étudiés, les primes totales de risque réagissent significativement, entre autres, aux événements suivants : les chocs pétroliers (1973, 1978), les réformes monétaires états-uniennes (1979-1982), le grand crash boursier (1987), les guerres du Golfe (1991,2003), la série de crises monétaires et financières des pays émergents asiatiques et latino-américains (1991,1993, 1997, 1998, 2001) et les attaques terroristes contre les Etats-Unis (2001). Ces résultats suggèrent que tous les marchés boursiers étudiés étaient exposés à ces événements internationaux. Nos résultats montrent aussi que pour le marché mondial et les marchés boursiers des pays développés considérés, ce sont les primes de risque du marché mondial qui traduisent le plus ces évènements internationaux. En revanche, pour les marchés boursiers des pays émergents étudiés (Singapour et Afrique du Sud), ce sont plutôt les primes liées au risque des taux de change qui réagissent le plus à ces évènements internationaux.

A présent nous allons étudier en détail les primes de risque marché par marché. Pour le marché français, la prime associée au risque des taux de change est relativement volatile et est en moyenne négative. Néanmoins, elle change considérablement de poids et de signe au cours du temps. Pour la période 1973-1975, la prime totale est pratiquement entièrement déterminée par la prime de risque de change. La prime de change devient plus faible durant la période 1976-1979. Pendant les années 1990, la prime de risque de change est relativement moins volatile que celle observée durant les années 1970 et 1980.

Le marché américain présente une prime de change significativement plus faible que le marché français et ce durant la période entière couverte par cette étude. Manifestement, la prime totale de risque est principalement déterminée par la prime de risque du portefeuille du marché mondial. Les mêmes observations restent valables pour le marché mondial. Ce résultat s'explique en partie par le poids important du marché américain dans le portefeuille du marché mondial.

L'inspection attentive des évolutions de la prime de risque pour le marché sud-africain montre l'importance du risque des taux de change. La prime de risque totale est principalement déterminée par la prime de risque de change et ce pour toute la période couverte par l'étude. Contrairement aux autres marchés boursiers discutés ci-dessus, le rôle de la prime de risque de change devient plus important durant les dernières années de notre échantillon.

Pour le marché boursier singapourien, cinq sous-périodes sont à distinguer. Pendant la période 1973-1979, la prime de risque de change est négative et sa contribution à la prime totale est significative. La prime de change devient positive et explique une bonne partie de la prime de risque totale pendant la sous-période 1980-1983. Durant la sous-période 1984-1996, la prime de risque des taux de change redevient de nouveau négative. Cependant, elle est relativement moins importante que pour les deux sous-périodes précédentes. Pour la sous-période 1997-1999, la prime de risque de change est négative et explique pratiquement la totalité de la prime totale de risque. Cette sous-période correspond à la série des crises financières et monétaires asiatiques. La dernière sous-période 2000-2003 est marquée par une prime de change relativement moins importante.

Le Tableau 4 présente pour chaque marché étudié des statistiques sur la prime de risque totale ainsi que sur sa décomposition en prime de risque du marché mondial et en prime de risque des taux de change. Cependant, il est utile de mentionner qu'étant donné les grandes fluctuations négatives et positives notamment des primes des risques des taux de change, il est difficile de fournir des statistiques ayant des valeurs informationnelles significatives. Afin de surmonter cette difficulté, les statistiques seront calculées pour la période entière couverte par l'étude 1973-2003, mais aussi pour les deux sous-périodes 1973-1989 et 1990-2003. De nombreuses études empiriques ont montré que les marchés financiers sont devenus plus intégrés dans les années 1990 qu'auparavant. Ainsi, le Tableau 4 permet aussi d'avoir une idée sur l'évolution des primes de risque en fonction de l'intégration financière.

Le Tableau 4 suscite plusieurs remarques. D'abord, excepté la prime totale de Singapour pour la seconde sous-période, les primes de risque sont toutes significativement différentes de zéro. Ensuite, excepté l'Afrique du Sud, les primes totales de risque ont diminué pour tous les marchés en allant de la sous-période 1970-1989 à la sous-période 1990-2003. Ces résultats suggèrent que les nouvelles opportunités d'investissement ont permis une meilleure diversification des risques et donc une baisse des prix et des quantités de risques, (voir Stulz (1999)).

Pendant la période entière couverte par l'étude 1973-2003, la prime associée au risque des taux de change est statistiquement et économiquement significative pour tous les marchés étudiés. Néanmoins, la contribution de la prime de change dans la prime totale est plus prononcée pour les marchés des pays émergents étudiés. Pour le marché mondial, la France, les Etats-Unis et Singapour, la prime de risque de change est négative. Cela signifie que les investisseurs de ces pays sont en moyenne prêts à payer une partie de leur prime totale pour se protéger contre les fluctuations non-anticipées des taux de change. Pour les marchés des pays développés étudiés (Etats-Unis et France), la prime de risque est déterminée en grande partie par la prime de risque du marché mondial. Pour Singapour, la prime de risque du portefeuille du marché mondial est relativement importante mais la prime totale est relativement faible, cela s'explique par une prime de risque des taux de change importante mais négative. Enfin, pour le marché sud-africain, la prime de risque de change contribue plus que la prime de risque du marché mondial à la formation de la prime totale. Le marché sud-africain a la plus grande prime de risque à cause de son importante prime de change positive.

La sous-période 1973-1989 est caractérisée par des primes de risque totales élevées traduisant ainsi les grandes turbulences qui l'ont marquée. Ces primes varient de 7.50% pour les Etats-Unis à 10.38% pour l'Afrique du Sud. Excepté pour l'Afrique du Sud, ces primes totales de risque sont déterminées en grande partie par l'exposition aux fluctuations du portefeuille du marché international. Ainsi, les primes de change enregistrées sont de –0.68%, –0.93%, -2.06% et –2.36% respectivement pour les Etats-Unis, le marché mondial, la France et Singapour. Pour l'Afrique du Sud, la prime de change contribue pour plus de 50% de la prime totale de risque.

Durant la sous-période 1990-2003, le rôle des primes de risque des taux de change devient plus important pour tous les marchés étudiés. Ceci semble refléter les séries de crises monétaires et financières qu'ont connues les marchés financiers durant cette période. Ainsi, ces primes de change passent à –1.41%, -1.75% et –2.11% respectivement pour les Etats-Unis, le monde et la France. L'augmentation du rôle négatif des primes des taux de change s'est accompagnée d'une baisse des primes de risque du marché mondial. Il s'ensuit que la prime totale a baissé significativement pour ces pays dans les dernières années. Ainsi, les primes totale passent de 8.02% à 5.16% pour le marché mondial, de 7.50% à 6.30% pour le marché américain et de 7.67% à 5.37% pour la France.

Pour Singapour, la seconde sous-période est marquée par une contribution très importante mais négative de la prime de change. Cette prime s'établit à –9.35%. La prime de risque du marché mondial a baissé légèrement pour s'établir à 7.97%. Il s'ensuit alors une baisse importante de la prime totale. Cette dernière chute de 8.18% à –1.39%. Toutefois, il est utile de noter que la prime totale est statistiquement non différente de zéro.

En outre, l'examen attentif des évolutions de la prime de risque singapourienne (voir graphique) montre que la prime de change a chuté brusquement pendant la période 1997-1998. Cette période correspond aux crises asiatiques. Pour la période 1997-1998, la prime de risque de change est de –31.30%, traduisant l'ampleur de la crise et les sacrifices en terme de prime de risque que les investisseurs étaient prêts à payer pour se couvrir contre les fluctuations non-anticipées des taux de change. Pendant la même période, la prime de marché était de 7.40%.

Pour le marché sud-africain, on a pu observer une baisse aiguë de la prime de risque du portefeuille mondial qui s'établit à seulement 2.81%, une hausse aussi aiguë de la prime de change qui passe à 10.51%, ce qui a fait augmenter significativement la prime totale à 13.39%.

En résumé, la seconde sous-période se caractérise par une contribution statistiquement et économiquement plus significative de la prime de change et par une baisse de la prime totale de risque pour tous les marchés étudiés à l'exception du marché sud-africain. Toutefois, il est utile de signaler que les primes de risque ont augmenté significativement pour tous les marchés considérés à partir de 2001, suite à l'augmentation de l'incertitude et de l'aversion au risque des investisseurs globaux dans les dernières années.

Pour conclure, la prime de change constitue une composante statistiquement et économiquement significative de la prime totale de risque. Ces résultats sont en accord avec ceux de De Santis et Gérard (1998) et De Santis et al. (2003). Plus particulièrement, nos résultats montrent que la contribution de la prime de change est plus prononcée pour les marchés des pays émergents étudiés que pour ceux des pays développés. Cependant, il est tout à fait raisonnable de se demander si l'importance de la prime de change ne pouvait tout simplement pas être due au fait que le modèle estimé ci-dessus suppose que les marchés étudiés sont parfaitement intégrés et donc n'inclut pas de facteurs locaux de risque. La théorie monétaire et financière nous apprend que les taux de change sont liés aux fondamentaux macroéconomiques des pays et donc une segmentation partielle peut être à l'origine des résultats trouvés. Un re-examen de la significativité de la prime de change dans le cadre plus réaliste de marchés financiers partiellement intégrés s'avère alors indispensable. Cet examen fera l'objet de la section qui suit.

### 5.2- Cas des marchés boursiers partiellement segmentés

Dans la sous-section 5.1, nous avons montré que sous l'hypothèse d'intégration financière parfaite la prime associée aux fluctuations des taux de change était statistiquement et économiquement significative pour l'ensemble des marchés étudiés. Toutefois, le fait d'imposer *a priori* l'hypothèse d'intégration financière parfaite peut relativiser l'étendu de ces résultats. En effet, les taux de change sont étroitement liés aux fondamentaux domestiques. Dans ce sens, le risque de change peut constituer un *proxy*, certes biaisé mais significatif, des sources locales de risque. Ainsi, il est raisonnable de penser que les primes de risque de change sont significatives tout simplement parce que le modèle estimé plus-haut ne considère pas les facteurs locaux de risque. Ces facteurs locaux de risque peuvent être non-diversifiables internationalement et donc rémunérés par des primes de risque. C'est notamment le cas de nombreux pays émergents ayant des marchés partiellement segmentés, (voir Bekaert et Harvey (1995)). Un re-examen de la significativité des primes des taux de change dans le cadre général de marchés boursiers partiellement segmentés s'avère alors nécessaire.

Le Tableau 5 résume les résultats de l'estimation d'un modèle qui suppose explicitement que les marchés sont caractérisés par une segmentation partielle. Ainsi, les rentabilités sont déterminées par une combinaison des facteurs locaux et globaux de risque. Précisément, la prime totale de risque est composée de cinq primes : une prime liée au risque du portefeuille du marché mondial, une prime liée au risque du portefeuille du marché domestique et trois primes associées aux fluctuations non-anticipées des taux de change des pays étudiés. Le système estimé est celui décrit par les relations (9) et (10). Afin de préserver de l'espace, seules les valeurs estimées des prix des risques domestiques sont reportées. Notons cependant que les structures des premiers et seconds moments conditionnels restent sensiblement inchangées par rapport à celles présentées dans le Tableau 3.

Les prix des risques domestiques résiduels ne sont pas individuellement significatifs aux seuils conventionnels de significativité de 1% et de 5%. Toutefois, le prix du risque domestique sud-africain est significatif à 10%. Ce résultat va en faveur d'une légère segmentation partielle caractérisant le marché sud-africain. Néanmoins, le test robuste de Wald ne permet pas de rejeter l'hypothèse de base selon laquelle les prix des risques domestiques sont conjointement nuls. Ces résultats suggèrent, du moins pour les marchés étudiés, que le risque domestique n'est pas un facteur de risque rémunéré internationalement pendant la période couverte par la présente étude.

Plus intéressant, les résultats relatifs aux prix et aux primes associés aux fluctuations non-anticipées des taux de change restent pratiquement inchangés. Les prix du risque de change sont significatifs pour tous les marchés étudiés. Les primes de risque des taux des changes sont statistiquement et économiquement significatives.

### 5.3- Tests de robustesse

Le Tableau 6 montre les résultats de quelques tests complémentaires de robustesse du modèle. En premier lieu, nous testons la sensibilité des résultats obtenus par rapport à la contrainte de non-négativité du prix de risque de marché mondial. En effet, dans le modèle discuté ci-dessus, une double hypothèse est imposée sur les signes des prix de risques : le prix de risque de marché est contraint à être non-négatif alors que les prix des risques des taux de changes sont autorisés à prendre tant des valeurs positives que des valeurs négatives. Nous estimons ainsi une nouvelle version du modèle où aussi bien le prix de risque de marché que les prix des risques des taux de change varient linéairement en fonction des instruments : $\delta_{m,t-1} = \kappa'_W Z_{t-1}$ et $\delta_{k,t-1} = \kappa'_k Z_{t-1}$. Nous re-testons alors la significativité de ces prix de risques. Les résultats de ces tests sont résumés dans le Panel A du Tableau 6. Toutes les hypothèses sont testées par le test robuste de Wald à partir des estimations du modèle par la méthode du quasi-maximum de vraisemblance.

Le prix du risque de marché mondial est significativement variable dans le temps. La Figure 1.b (voir annexe) montre la dynamique comparée des prix de risque du marché mondial estimés avec et sans la contrainte de non-négativité (*i.e.* fonction exponentielle versus fonction linéaire). Les deux séries présentent une corrélation de 78%. Néanmoins, des différences significatives existent entre les dynamiques des deux séries. En particulier, la série du prix du risque linéaire est caractérisée par des périodes de prix de risque négatif. C'est notamment le cas pour la sous-période 1978-1982. Boudoukh et al. (1993) trouve pour cette période des primes de risque négatives pour le marché américain. Les auteurs expliquent ce résultat par une forte inflation et une structure inversée des taux d'intérêt.

Ce qui nous intérese le plus dans cet exercice c'est la significativité des prix des risques des taux de change. L'hypothèse selon laquelle le prix de risque de change est nul est rejetée à 1% pour la France et Singapour et à 5% pour l'Afrique du Sud. L'hypothèse selon laquelle le prix de change est constant est rejetée à 1% pour la France, à 5% pour Singapour et à 10% pour l'Afrique du Sud. L'hypothèse selon laquelle les prix des risques des taux de change sont conjointement égaux à zéro est rejetée à tous les seuils habituels de significativité. De même, on rejette fortement l'hypothèse selon laquelle les prix des risques des taux de change sont conjointement constants. Ces résultats confirment ceux obtenus pour la spécification avec contrainte de non-négativité et montrent que le risque des taux de change est significativement rémunéré aussi bien pour les marchés des pays développés que pour ceux des pays émergents étudiés. Ainsi, les résultats discutés dans la section précédente paraissent peu sensibles à la contrainte de non-négativité imposée sur le prix de risque de marché mondial.

En second lieu, une version augmentée du modèle (relation (12)) est estimée simultanément pour tous les marchés étudiés. D'abord, un terme constant ($\alpha_i$), spécifique à chaque pays, est introduit. Ce terme constant est censé capturer les autres formes de segmentation non intégrées dans le modèle de base. Il peut par exemple s'agir des coûts d'information et de transaction, de traitement fiscal discriminatoire ou d'autres manifestations de souverainetés nationales. Si le MEDAFI est valide, les termes constants sont individuellement et conjointement non-significativement différents de zéro. En outre, dans la version augmentée du modèle, on introduit également les variables d'information globales utilisées afin de conditionner l'estimation du modèle. L'introduction de ces variables peut s'interpréter comme un test de leur pouvoir prédictif une fois que ces variables d'information sont utilisées pour conditionner les estimations. Si notre modèle est bien spécifié, ces variables instrumentales n'ont aucun pouvoir prédictif.

$$E(R_{i,t}^c / \Omega_{t-1}) - R_{f,t}^c = \alpha_i + \delta_{m,t-1} Cov(R_{i,t}^c, R_{m,t}^c / \Omega_{t-1}) + \sum_{k=1}^{3} \delta_{k,t-1} Cov(R_{i,t}^c, R_{k,t}^c / \Omega_{t-1}) \\ + \delta_{di} Var(Res_{it}^c / \Omega_{t-1}) + \varphi_k' Z_{t-1} ; \quad \forall i \qquad (12)$$

Les résultats de cet exercice sont résumés dans le Panel B du Tableau 6. Les termes constants sont significativement conjointement nuls. Les barrières directes de type coûts de transaction et d'information ne constituent pas une source significative de segmentation pour les marchés étudiés. Ce résultat semble être confirmé par les prix des risques domestiques qui ne sont pas conjointement statistiquement différents de zéro. En outre, nos résultats montrent que les variables d'information n'ont plus de pourvoir prédictif une fois qu'elles sont utilisées pour conditionner les estimations.

## 6- Conclusion

L'objet de cet article était d'étudier les déterminants et la dynamique de la prime de risque dans un environnement international. Pour ce faire, nous avons développé une version conditionnelle multivariée du modèle international d'évaluation des actifs financiers adaptée à la fois aux marchés développés et aux marchés émergents. Cette version permet de rendre compte des déviations de la PPA et des différentes structures des marchés boursiers. Contrairement aux études antérieures, nous n'avons pas imposé l'hypothèse restrictive d'inflation locale non-stochastique ou nulle. Plus précisément, nous avons utilisé les taux de change réels, ce qui permet de surmonter les difficultés associées à l'estimation du modèle MEDAFI de Adler et Dumas (1983) dans le cas des pays dont les taux d'inflation sont volatiles.

Le MEDAFI proposé a ensuite été estimé pour la période 1973-2003 conjointement pour cinq marchés (deux marchés développés (France et Etats-Unis), deux marchés émergents (Afrique du Sud et Singapour) plus le marché mondial), et ce en utilisant le processus GARCH multivarié de De Santis et Gérard (1997). Une attention particulière a été accordée au risque des taux de change et aux évolutions des différentes composantes de la prime de risque.

Les résultats de l'estimation du modèle MEDAFI avec déviations de la PPA sous l'hypothèse d'intégration financière parfaite ont montré que les prix des risques des taux de change sont significatifs et variables dans le temps aussi bien pour les marchés des pays développés que pour ceux des pays émergents considérés. La décomposition de la prime de risque totale en prime de change et en prime de marché a montré que cette dernière constitue souvent la composante principale de la prime totale. Néanmoins, la contribution des primes de change est économiquement et statistiquement significative pour tous les marchés étudiés. Les primes de change varient considérablement dans le temps et d'un marché à l'autre. En particulier, la contribution des primes de change à la formation de la prime totale de risque est plus importante pour les marchés des pays émergents que pour ceux des marchés développés considérés. En outre, nos résultats ont montré qu'à l'exception de l'Afrique du Sud, la prime de risque totale a diminué considérablement dans les dernières années pour tous les marchés étudiés.

Toutefois, la théorie économique nous enseigne que les taux de change sont liés aux fondamentaux nationaux. Dès lors, nous nous sommes demandés si la significativité des primes de change n'était tout simplement pas due au fait que le modèle estimé sous l'hypothèse d'intégration financière parfaite ignorait les sources domestiques de risque. Ainsi, l'extension du modèle au cas plus général de marchés financiers partiellement segmentés s'est avérée nécessaire, surtout si l'on prend en considération que l'étude porte aussi bien sur des marchés développés que sur des marchés émergents.

L'estimation du MEDAFI avec déviations de la PPA et segmentation partielle a montré qu'excepté pour l'Afrique du Sud, les prix des risques domestiques ne sont pas individuellement significatifs et que globalement on ne peut pas rejeter l'hypothèse d'intégration financière. En particulier, les résultats obtenus relatifs aux prix et aux primes de change restent pratiquement inchangés par rapport à ceux du modèle estimé sous l'hypothèse d'intégration financière parfaite. Les prix du risque de change sont significatifs pour tous les marchés étudiés et les primes de risque de change contribuent significativement à la formation de la prime totale de risque.

**Tableau 1**
*Statistiques Descriptives des séries de Rentabilités*

*Les rentabilités boursières sont exprimées en dollar américain et calculées avec réinvestissement des dividendes en excès du taux des eurodollars à 30 jours. Les colonnes TcrFra, TcrSinga et TcrAfri sont réservées aux rentabilités des taux de change réels. Toutes les rentabilités sont mensuelles et couvrent la période février 1973- mai 2003.*

**Panel A: Statistiques descriptives**

|  | France | Singa. | Afri. Sud | Etats-Unis | TcrFra | TcrSinga | TcrAfri | Monde |
|---|---|---|---|---|---|---|---|---|
| **Moyenne (% par année)** | 6.22 | 3.28 | 9.28 | 4.93 | 1.32 | -1.36 | 7.74 | 3.73 |
| **Ecart-type (% par année)** | 81.95 | 103.67 | 100.77 | 55.66 | 35.15 | 16.82 | 39.26 | 52.03 |
| **Skewness** | -0.02 | 0.31** | -0.37* | -0.28** | -1.43* | -0.12 | 1.05* | -0.37* |
| **Kurtosis**[1] | 1.35* | 5.33* | 1.45* | 1.69* | 11.84* | 3.59* | 8.62* | 1.17* |
| **J.B.** | 27.54* | 435.42* | 40.57* | 48.18* | 2241.59* | 195.64* | 1188.92* | 29.28* |
| **Q(12)** | 10.95 | 16.89 | 7.21 | 11.78 | 28.46* | 31.53* | 97.62* | 15.14 |

**Panel B: Corrélations non-conditionnelles des $r_{it}$**

|  | France | Singa. | Afri. Sud | Etats-Unis | TcrFra | TcrSinga | TcrAfri | Monde |
|---|---|---|---|---|---|---|---|---|
| **France** | 1.00 | 0.35 | 0.39 | 0.52 | -0.24 | -0.20 | -0.08 | 0.69 |
| **Singa.** |  | 1.00 | 0.30 | 0.52 | -0.18 | -0.33 | -0.06 | 0.59 |
| **Afri. Sud** |  |  | 1.00 | 0.34 | -0.17 | -0.23 | -0.28 | 0.46 |
| **Etats-Unis** |  |  |  | 1.00 | -0.07 | -0.11 | -0.18 | 0.85 |
| **TcrFra** |  |  |  |  | 1.00 | 0.50 | 0.25 | -0.18 |
| **TcrSinga** |  |  |  |  |  | 1.00 | 0.27 | -0.22 |
| **TcrAfri** |  |  |  |  |  |  | 1.00 | -0.08 |
| **Monde** |  |  |  |  |  |  |  | 1.00 |

**Panel C: Autocorrelations des $r_{it}$**

| Retard | France | Singa. | Afri. Sud | Etats-Unis | TcrFra | TcrSinga | TcrAfri | Monde |
|---|---|---|---|---|---|---|---|---|
| 1 | 0.042 | 0.067 | 0.072 | -0.007 | 0.255* | 0.260* | 0.372* | 0.053 |
| 2 | -0.025 | -0.001 | -0.027 | -0.035 | 0.019 | 0.020 | 0.148* | -0.049 |
| 3 | 0.098 | -0.0056 | -0.002 | 0.034 | 0.070 | -0.051 | 0.034 | 0.045 |
| 4 | 0.029 | 0.036 | -0.045 | -0.032 | 0.009 | -0.065 | -0.104 | -0.017 |
| 5 | 0.026 | -0.039 | -0.071 | 0.113 | 0.036 | -0.082 | -0.093 | 0.092 |
| 6 | 0.042 | -0.109 | 0.002 | -0.039 | -0.015 | -0.037 | -0.064 | -0.036 |

**Panel D: Autocorrelations des $(r_{it})^2$**

| Retard | France | Singa. | Afri. Sud | Etats-Unis | TcrFra | TcrSinga | TcrAfri | Monde |
|---|---|---|---|---|---|---|---|---|
| 1 | 0.070 | 0.129** | 0.051 | 0.118** | -0.015 | 0.265* | 0.167* | 0.049 |
| 2 | 0.010 | 0.042 | 0.283* | 0.060 | -0.019 | 0.157* | 0.008 | 0.044 |
| 3 | 0.041 | 0.050 | 0.045 | 0.114** | -0.015 | 0.155* | 0.094 | 0.023 |
| 4 | 0.142* | 0.058 | 0.078 | 0.081 | -0.002 | 0.152* | 0.112 | 0.019 |
| 5 | 0.017 | 0.036 | 0.081 | -0.001 | 0.039 | 0.184* | 0.235* | 0.069 |
| 6 | 0.045 | 0.071 | 0.016 | 0.027 | 0.011 | 0.93 | 0.075 | 0.040 |

**Panel E: Corrélations croisées des $(r_{it})^2$ - Marché mondial et pays i**

| Retard | France | Singa. | Afri. Sud | Etats-Unis | TcrFra | TcrSinga | TcrAfri |
|---|---|---|---|---|---|---|---|
| -6 | 0.000 | 0.073 | 0.017 | 0.040 | -0.059 | 0.165* | -0.017 |
| -5 | 0.043 | 0.116** | -0.002 | 0.042 | 0.015 | 0.096 | -0.010 |
| -4 | 0.128** | 0.098 | -0.055 | -0.031 | 0.096 | 0.008 | -0.003 |
| -3 | 0.061 | 0.076 | -0.007 | 0.046 | -0.052 | 0.021 | 0.060 |
| -2 | -0.001 | 0.032 | 0.093 | 0.085 | 0.0516 | 0.168* | -0.010 |
| -1 | -0.012 | 0.031 | -0.023 | 0.078 | -0.014 | 0.041 | -0.010 |
| 0 | 0.517* | 0.585* | 0.350* | 0.805* | 0.003 | 0.196* | -0.028 |
| 1 | 0.072 | 0.060 | 0.051 | 0.030 | -0.030 | -0.001 | 0.111** |
| 2 | -0.040 | 0.007 | 0.090 | 0.019 | -0.010 | 0.065 | 0.016 |
| 3 | -0.061 | 0.055 | 0.052 | 0.136** | -0.009 | -0.003 | 0.009 |
| 4 | 0.093 | -0.035 | 0.037 | 0.072 | -0.021 | -0.023 | 0.023 |
| 5 | 0.011 | -0.027 | 0.007 | 0.004 | 0.023 | 0.061 | -0.051 |
| 6 | 0.033 | 0.099 | -0.010 | 0.028 | 0.079 | 0.065 | 0.046 |

*Nombre des corrélations croisées d'ordre (-2, -1,1,2) significatives: 11 sur 112.*

\* significatif à 1%, \*\* significatif à 5%, $r_{it}$ série de rentabilités, (1) centré sur 3, Q(12) :test de Ljung-Box d'ordre 12 et J.B. test de normalité de Jarque-Bera

**Tableau 2**
*Statistiques descriptives et corrélations des variables d'information*

*Le vecteur $Z_{t-1}$ de variables d'information internationales contient: un terme constant, le rendement en dividende (dividend price ratio) du portefeuille de marché mondial en excès du taux des eurodollars à 30 jours (RDM), la variation mensuelle d'une prime de terme américaine (DPTEU), une prime de défaut américaine (PDEU) et la variation mensuelle du rendement d'un certificat américain de trésorerie à 30 jours (DTIM).*

**Panel A: Statistiques descriptives**

|  | RDM | DPTEU | PDEU | DTIM |
|---|---|---|---|---|
| **Moyenne** | -0.28 | 0.01 | 1.10 | -0.01 |
| **Ecart-type** | 0.187 | 0.48 | 0.45 | 0.68 |
| **Skewness** | -0.69* | 0.71 * | 1.07* | -1.94* |
| **Kurtosis**[1] | 1.36* | 12.51 * | 0.90* | 29.72* |
| **J.B.** | 57.74* | 2394.14* | 82.28* | 13553.34* |
| **Q(12)** | 2252.23* | 41.37* | 2653.39* | 128.87* |

**Panel B: Corrélations**

|  | RDM | DPTEU | PDEU | DTIM |
|---|---|---|---|---|
| **RDM** | 1.00 | 0.12 | -0.28 | -0.17 |
| **DPTEU** |  | 1.00 | 0.13 | -0.58 |
| **PDEU** |  |  | 1.00 | -0.17 |
| **DTIM** |  |  |  | 1.00 |

*\* significatif à 1%, \*\* significatif à 5%, (1) centré sur 3, Q(12) :test de Ljung-Box d'ordre 12 et J.B. :test de normalité de Jarque-Bera*

**Tableau 3**
*Estimation par la méthode du quasi-maximum de vraisemblance du MEDAFI conditionnel*

*Le modèle estimé est le suivant :*

$$r_t = \delta_{m,t-1} h_{m,t} + \sum_{k=1}^{3} \delta_{k,t-1} h_{4+k,t} + \varepsilon_t \quad , \quad \varepsilon_t / \Omega_{t-1} \sim N(0, H_t)$$

$$\delta_{m,t-1} = \exp(\kappa'_W Z_{t-1}) \quad , \quad \delta_{k,t-1} = \kappa'_k Z_{t-1} \quad , \quad k = 1,2,3 \; ;$$

$$H_t = H_0 * (\tau\tau' - aa' - bb') + aa' * \varepsilon_{t-1}\varepsilon'_{t-1} + bb' * H_{t-1}$$

*où $r_t$ est le vecteur ($8 \times 1$) de rentabilités, $\varepsilon_t$ est le vecteur ($8 \times 1$) des résidus, $H_t$ est la matrice ($8 \times 8$) des variances-covariances conditionnelles, $h_{4+k,t}$ et $h_{m,t}$ sont respectivement la $4+k^{ème}$ et la dernière colonne de $H_t$, $H_0$ est la matrice des variances-covariances non-conditionnelles, **a** et **b** sont deux vecteurs ($8 \times 1$) de paramètres inconnus, $\tau$ est le vecteur unitaire de taille ($8 \times 1$) et \* le produit matriciel de Hadamard.*

**Panel A: Prix des risques**

*(a) Prix du risque du marché mondial*

|  | Const. | RDM | DPTEU | PDEU | DTIM |
|---|---|---|---|---|---|
| **Prix du risque du marché mondial** | 0.904* | 0.120* | -0.928* | 0.491** | -0.815* |
|  | (0.001) | (0.028) | (0.242) | (0.224) | (0.301) |

*(b) Prix des risques des taux de change*

|  | Const. | RDM | DPTEU | PDEU | DTIM |
|---|---|---|---|---|---|
| **Prix du risque du franc français** | -1.428 | -20.234** | 28.401** | -1.914 | 13.230* |
|  | (1.142) | (7.256) | (12.77) | (1.112) | (3.641) |
| **Prix du risque du dollar singapourien** | 27.923 | 7.557** | -14.082* | -14.093** | -11.657* |
|  | (18.775) | (4.257) | (6.323) | (8.087) | (3.905) |
| **Prix du risque du rand sud-africain** | -18.959* | 1.636 | -16.091** | 9.015 | -12.350** |
|  | (7.263) | (18.563) | (7.684) | (6.571) | (6.680) |

**Panel B: Processus GARCH**

| | France | Singapour | Afri. Sud | Etats-Unis | TcrFra | TcrSing | TcrAfri | Monde |
|---|---|---|---|---|---|---|---|---|
| *a* | 0.093* | 0.355* | 0.268* | 0.036* | 0.191* | 0.431* | 0.370* | 0.170* |
| | (0.007) | (0.036) | (0.043) | (0.007) | (0.067) | (0.035) | (0.014) | (0.027) |
| *b* | 0.995* | 0.806* | 0.895* | 0.982* | 0.762* | 0.836* | 0.791* | 0.348** |
| | (0.002) | (0.044) | (0.040) | (0.002) | (0.166) | (0.032) | (0.011) | (0.152) |

**Panel C: Tests de specification**

| Hypothèse nulle | $\chi^2$ | df | p-value |
|---|---|---|---|
| *Le prix mondial du risque est-il constant ?* $H_0: \delta_{m,j}=0 \quad \forall j>1$ | 207.826 | 4 | 0.000 |
| *Le prix du risque du franc français est-il égal à zéro?* $H_0: \delta_{k,j}=0 \; ; k=1 \quad \forall j$ | 16.971 | 5 | 0.005 |
| *Le prix du risque du franc français est-il constant?* $H_0: \delta_{k,j}=0 \; ; k=1 \quad \forall j>1$ | 16.971 | 4 | 0.002 |
| *Le prix du risque du dollar singapourien est-il égal à zéro?* $H_0: \delta_{k,j}=0 \; ; k=2 \quad \forall j$ | 21.025 | 5 | 0.001 |
| *Le prix du risque du dollar singapourien est-il constant?* $H_0: \delta_{k,j}=0 \; ; k=2 \quad \forall j>1$ | 20.924 | 4 | 0.000 |
| *Le prix du risque du rand sud-africain est-il égal à zéro?* $H_0: \delta_{k,j}=0 \; ; k=3 \quad \forall j$ | 14.904 | 5 | 0.011 |
| *Le prix du risque du rand sud-africain est-il constant?* $H_0: \delta_{k,j}=0 \; ; k=3 \quad \forall j>1$ | 7.838 | 4 | 0.097 |
| *Les prix des risques des taux de change sont-ils conjointement nuls?* $H_0: \delta_{k,j}=0 \quad \forall j,k$ | 88.924 | 15 | 0.000 |
| *Les prix des risques des taux de change sont-ils conjointement constants?* $H_0: \delta_{k,j}=0 \quad \forall k \quad \forall j>1$ | 74.872 | 12 | 0.000 |

**Panel D: Diagnostic des résidus**

| | France | Singa. | Afri. Sud | Etats-Unis | TcrFra | TcrSinga | TcrAfri | Monde |
|---|---|---|---|---|---|---|---|---|
| **Skewness** | 0.01 | 0.36 | -0.30** | -0.29** | -1.42* | -0.11 | 1.06* | -0.38* |
| **Kurtosis** | 1.12* | 5.08* | 1.24* | 1.61* | 11.75* | 3.60* | 8.70* | 1.12* |
| **J.B.** | 18.95* | 398.47* | 29.04* | 44.56* | 2205.07* | 197.20* | 1210.58* | 28.13* |
| **Q(12)** | 12.20 | 17.57 | 6.65 | 12.90 | 27.83* | 30.93* | 96.48* | 17.49 |

*\* significatif à 1%, \*\* significatif à 5%, (1) centré sur 3, Q(12) :test de Ljung-Box d'ordre 12 et J.B. :test de normalité de Jarque-Bera. L'écart-type est reporté entre parenthèses.*

**Tableau 4**
*Décomposition des primes totales de risque*

*Le tableau contient les moyennes annualisées en pourcentage des primes de risque calculées à partir des résultats de l'estimation du modèle reportés dans le tableau 3. La prime totale (PT), la prime de marché (PRM) et la prime de change (PRC) sont mesurées comme suit:*

$$PRM_{it} = \delta_{m,t-1} Cov(r_{it}, r_{mt} / \Omega_{t-1})$$

$$PRC_{it} = \sum_{k=1}^{3} \delta_{k,t-1} Cov(r_{it}, r_{4+k,t} / \Omega_{t-1})$$

$$PT_{it} = \delta_{m,t-1} Cov(r_{it}, r_{mt} / Z_{t-1}) + \sum_{k=1}^{3} \delta_{k,t-1} Cov(r_{it}, r_{4+k,t} / \Omega_{t-1})$$

|  | PT | PRM | PRC | PT | PRM | PRC | PT | PRM | PRC |
|---|---|---|---|---|---|---|---|---|---|
|  | *1973-2003* | | | *1973-1989* | | | *1990-2003* | | |
| **Monde** | 6.769* | 8.056* | -1.287* | 8.017* | 8.944* | -0.927* | 5.163* | 6.609* | -1.745* |
|  | (0.417) | (0.381) | (0.066) | (0.708) | (0.661) | (0.092) | (0.258) | (0.192) | (0.077) |
| **Etats-Unis** | 6.358* | 7.359* | -1.001* | 7.500* | 8.180* | -0.680* | 4.886* | 6.297* | -1.411* |
|  | (0.370) | (0.348) | (0.059) | (0.631) | (0.605) | (0.087) | (0.217) | (0.169) | (0.063) |
| **France** | 6.673* | 8.748* | -2.075* | 7.665* | 9.729* | -2.064* | 5.368* | 7.479* | -2.110* |
|  | (0.597) | (0.417) | (0.283) | (1.020) | (0.724) | (0.478) | (0.402) | (0.207) | (0.227) |
| **Singapour** | 3.967* | 9.426* | -5.459* | 8.179* | 10.555* | -2.355* | -1.385 | 7.965* | -9.349* |
|  | (0.749) | (0.494) | (0.579) | (0.887) | (0.854) | (0.464) | (1.135) | (0.240) | (1.096) |
| **S. Africa** | 11.755* | 4.039* | 7.715* | 10.380* | 4.971* | 5.409* | 13.391* | 2.881* | 10.510* |
|  | (0.531) | (0.261) | (0.535) | (0.722) | (0.442) | (0.712) | (0.782) | (0.178) | (0.772) |

*\* significatif à 1%, \*\* significatif à 5%. L'écart-type est reporté entre parenthèses.*

**Tableau 5**
*MEDAFI conditionnel à Intégration Partielle*

*Le modèle estimé est le suivant :*

$$r_t = \delta_{m,t-1} h_{m,t} + \sum_{k=1}^{3} \delta_{k,t-1} h_{4+k,t} + \delta_d * q_t + \varepsilon_t \quad , \quad \varepsilon_t / \Omega_{t-1} \sim N(0, H_t)$$

$$\delta_{m,t-1} = \exp(\kappa'_W Z_{t-1}) \quad , \quad \delta_{k,t-1} = \kappa'_k Z_{t-1} \quad , \quad k=1,2,3 \;;$$

$$q_t = D(H_t) - (h_{mt} * h_{mt}) / h_{mmt}$$

$$H_t = H_0 * (\tau\tau' - aa' - bb') + aa' * \varepsilon_{t-1}\varepsilon'_{t-1} + bb' * H_{t-1}$$

*où $r_t$ est le vecteur ($8\times1$) de rentabilités, $\varepsilon_t$ est le vecteur ($8\times1$) des résidus, $H_t$ est la matrice ($8\times8$) des variances-covariances conditionnelles, $h_{4+k,t}$ et $h_{m,t}$ sont respectivement la $4+k^{ème}$ et la dernière colonne de $H_t$, $h_{mmt}$ est la variance du portefeuille du marché mondial, $D(H_t)$ est la diagonale de la matrice $H_t$, $\delta_d$ est le vecteur des prix des risques domestiques, $H_0$ est la matrice des variances-covariances non-conditionnelles, **a** et **b** sont deux vecteurs ($8\times1$) de paramètres inconnus, $\tau$ est le vecteur unitaire de taille ($8\times1$) et \* le produit matriciel de Hadamard.*

**Panel A: Tests de nullité individuelle**

|  | Etats-Unis | Singapour | France | Afrique du Sud |
|---|---|---|---|---|
| *Prix du risque domestique* | 0.161 | 1.827 | 1.253 | 3.144 |
|  | (0.933) | (1.879) | (1.662) | (1.908) |

**Panel B: Test de nullité jointe**

|  | $\chi^2$ | *df* | *p-value* |
|---|---|---|---|
| *Les prix de risque spécifiques sont-ils conjointement nuls?* <br> $H_0: \delta_{di} = 0 \quad \forall i$ | 4.196 | 4 | 0.380 |

*\* significatif à 1%, \*\* significatif à 5%. L'écart-type est reporté entre parenthèses.*

**Tableau 6**
**Tests de robustesse**

*Les modèles testés sont les suivants :*
***Panel A : Modèle avec prix de risque linéaires***

$$r_t = \delta_{m,t-1}h_{m,t} + \sum_{k=1}^{3} \delta_{k,t-1}h_{4+k,t} + \varepsilon_t \quad , \quad \varepsilon_t = (\varepsilon_1, \varepsilon_2, ..., \varepsilon_8)' / \Omega_{t-1} \sim N(0, H_t)$$

$$\delta_{m,t-1} = \kappa'_W Z_{t-1} \quad , \quad \delta_{k,t-1} = \kappa'_k Z_{t-1} \quad , \quad k=1,2,3 \ ;$$

***Panel B : Version augmentée du modèle***

$$R^c_{i,t} - R^c_{f,t} = \alpha_i + \delta_{m,t-1}Cov(R^c_{i,t}, R^c_{m,t}/\Omega_{t-1}) + \sum_{k=1}^{3}\delta_{k,t-1}Cov(R^c_{i,t}, R^c_{k,t}/\Omega_{t-1}) + \delta_{di}Var(Res^c_{it}/\Omega_{t-1}) + \varphi'_{ik}Z_{t-1} + \varepsilon_i \ ; \quad \forall i$$

$$\delta_{m,t-1} = \exp(\kappa'_W Z_{t-1}) \quad , \quad \delta_{k,t-1} = \kappa'_k Z_{t-1} \quad , \quad k=1,2,3 \ ;$$

$$Var(Res^c_{it}/\Omega_{t-1}) = Var(R^c_{it}/\Omega_{t-1}) - Cov(R^c_{it}, R^c_{mt}/\Omega_{t-1})^2 / Var(R^c_{mt}/\Omega_{t-1})$$

$$H_t = H_0 * (\tau\tau' - aa' - bb') + aa' * \varepsilon_{t-1}\varepsilon'_{t-1} + bb' * H_{t-1}$$

où $r_t$ est le vecteur $(8\times 1)$ de rentabilités, $\varepsilon_t$ est le vecteur $(8\times 1)$ des résidus, $H_t$ est la matrice $(8\times 8)$ des variances-covariances conditionnelles, $h_{4+k,t}$ et $h_{m,t}$ sont respectivement la $4+k^{ème}$ et la dernière colonne de $H_t$, $R^c_i$ est la rentabilité du marché $i$, $R^c_f$ est la rentabilité de l'actif sans risque, $\delta_{di}$ est le prix du risque domestique du marché $i$, $H_0$ est la matrice des variances-covariances non-conditionnelles, **a** et **b** sont deux vecteurs $(8\times 1)$ de paramètres inconnus, $\tau$ est le vecteur unitaire de taille $(8\times 1)$ et * le produit matriciel de Hadamard.

**Panel A : Modèle avec prix de risque linéaires**

| Hypothèse nulle | $\chi^2$ | df | p-value |
|---|---|---|---|
| *Le prix mondial du risque est-il constant ?* <br> $H_0: \delta_{m,j}=0 \quad \forall j>1$ | 259.910 | 4 | 0.000 |
| *Le prix du risque du franc français est-il égal à zéro?* <br> $H_0: \delta_{k,j}=0 \ ; k=1 \quad \forall j$ | 18.174 | 5 | 0.002 |
| *Le prix du risque du franc français est-il constant?* <br> $H_0: \delta_{k,j}=0 \ ; k=1 \quad \forall j>1$ | 15.851 | 4 | 0.003 |
| *Le prix du risque du dollar singapourien est-il égal à zéro?* <br> $H_0: \delta_{k,j}=0 \ ; k=2 \quad \forall j$ | 30.328 | 5 | 0.000 |
| *Le prix du risque du dollar singapourien est-il constant?* <br> $H_0: \delta_{k,j}=0 \ ; k=2 \quad \forall j>1$ | 11.410 | 4 | 0.022 |
| *Le prix du risque du rand sud-africain est-il égal à zéro?* <br> $H_0: \delta_{k,j}=0 \ ; k=3 \quad \forall j$ | 11.473 | 5 | 0.042 |
| *Le prix du risque du rand sud-africain est-il constant?* <br> $H_0: \delta_{k,j}=0 \ ; k=3 \quad \forall j>1$ | 8.172 | 4 | 0.085 |
| *Les prix des risques des taux de change sont-ils conjointement nuls?* <br> $H_0: \delta_{k,j}=0 \quad \forall j,k$ | 72.242 | 15 | 0.000 |
| *Les prix des risques des taux de change sont-ils conjointement constants?* <br> $H_0: \delta_{k,j}=0 \quad \forall k \quad \forall j>1$ | 51.120 | 12 | 0.000 |

**Panel B : Version augmentée du modèle**

| Hypothèse nulle | $\chi^2$ | df | p-value |
|---|---|---|---|
| *Les $\alpha_i$ sont-ils conjointement nuls?* <br> $H_0: \alpha_i=0 \quad \forall i$ | 3.496 | 4 | 0.476 |
| *Les prix de risque spécifiques sont-ils conjointement nuls?* <br> $H_0: \delta_{di}=0 \quad \forall i$ | 5.979 | 4 | 0.201 |
| *Les variables d'information sont-elles orthogonales aux rentabilités ?* <br> $H_0: \varphi_k=0 \quad \forall k$ | 4.551 | 4 | 0.336 |

## *Figure 1*
**Prix de risque de marché**

*1.a- Prix de risque de marché : fonction exponentielle*       *1.b- Prix de risque de marché : fonction linéaire versus exponentielle*

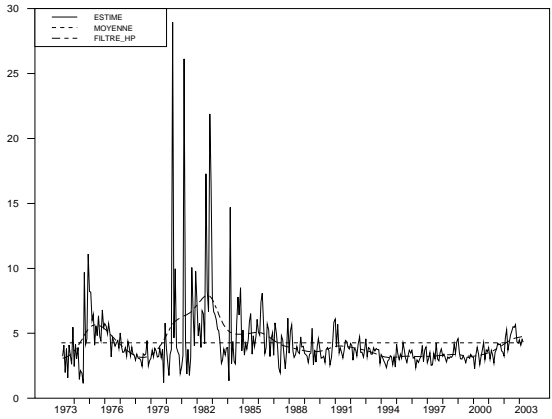 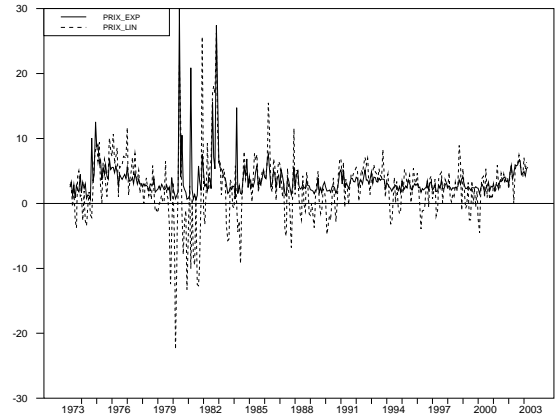

## *Figure 2*
**Prime de risque: France**

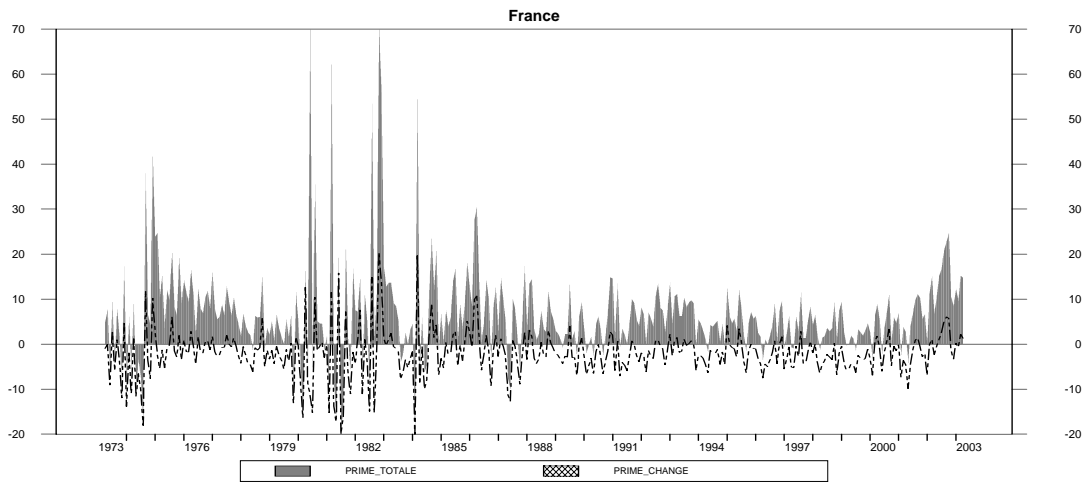

## *Figure 3*
**Prime de risque : Etats-Unis**

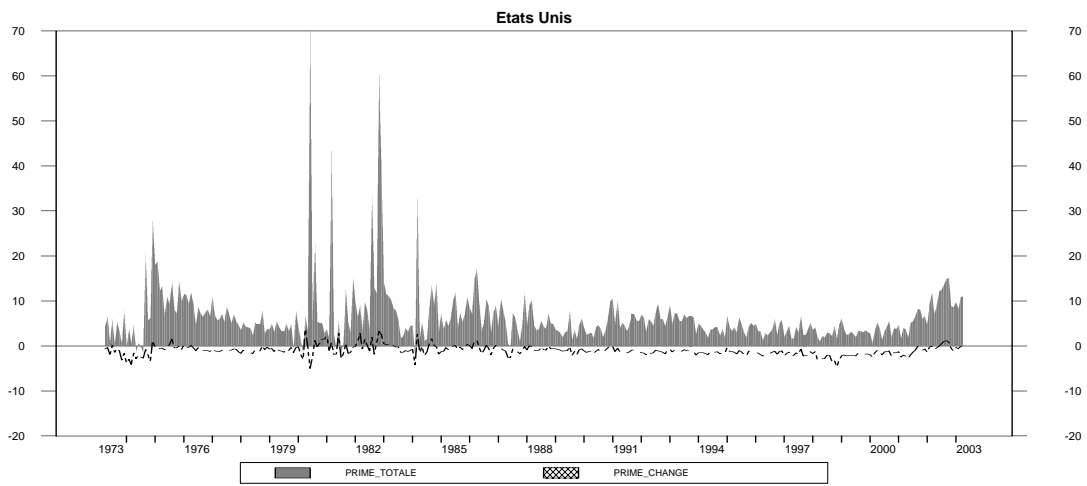

*Figure 4*
**Prime de risque : Singapour**

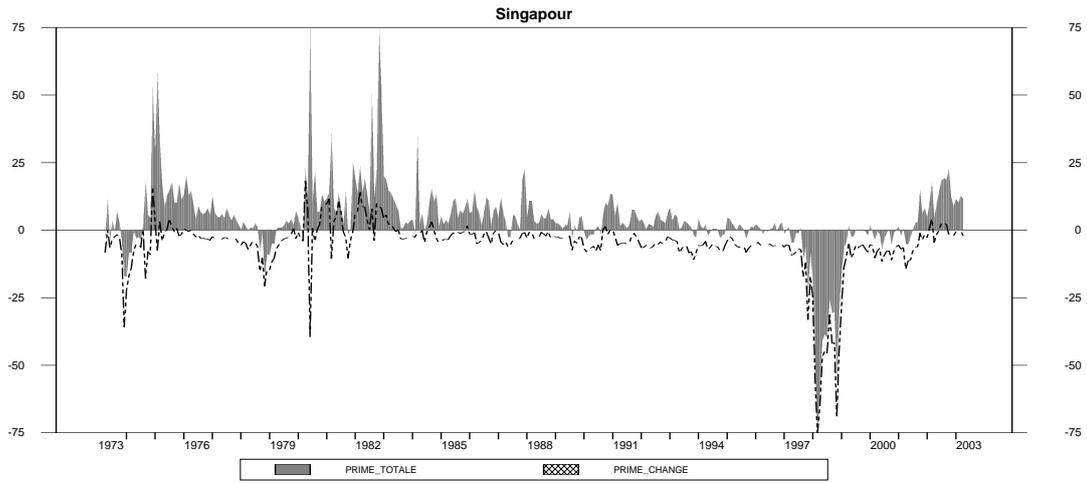

*Figure 5*
**Prime de risque : Afrique du Sud**

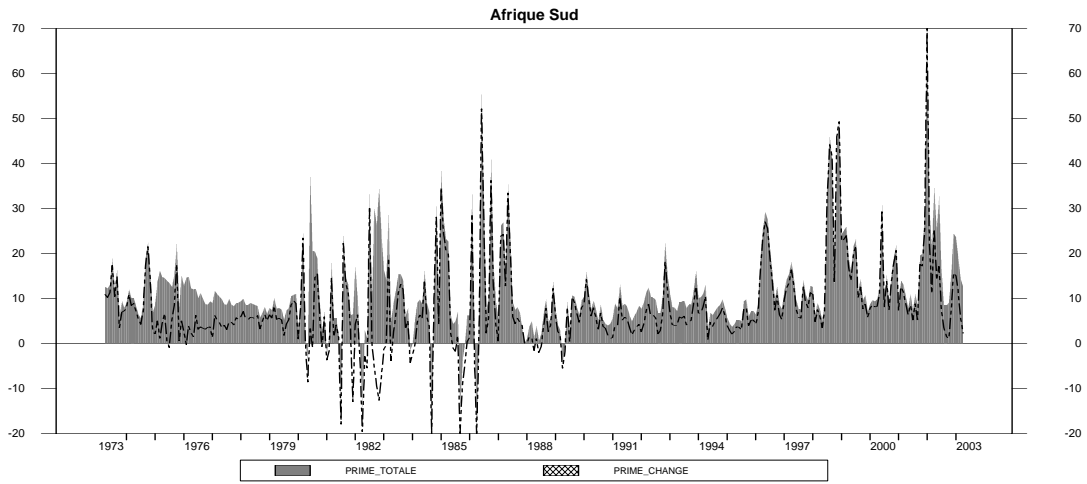

*Figure 6*
**Prime de risque : Monde**

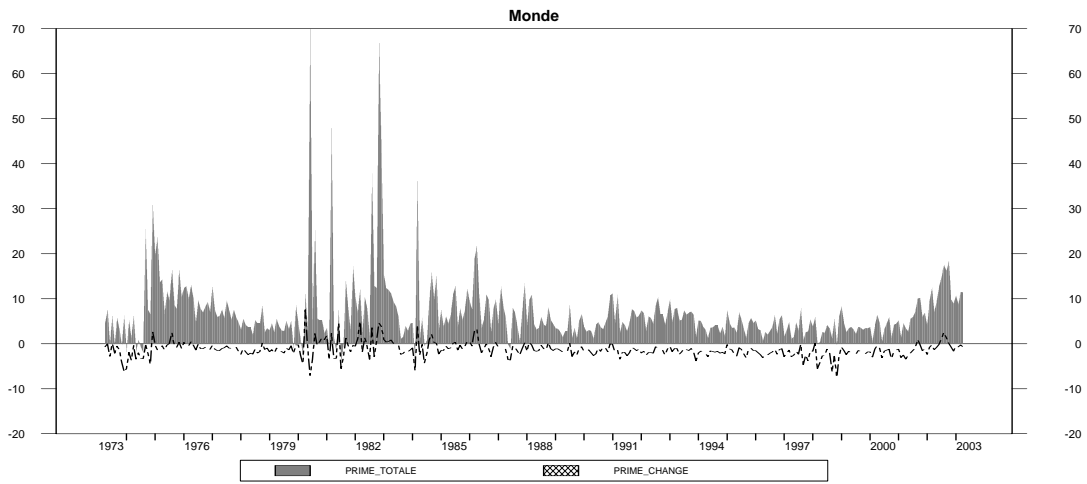

27